\documentclass[aps,prl,twocolumn,superscriptaddress,showpacs,floatfix]{revtex4-1}

\usepackage[utf8]{inputenc}
\usepackage[T1]{fontenc}
\usepackage{graphicx,color,amsmath,amsfonts,amssymb,xspace,epsfig,float,multirow}
\usepackage{lipsum}
\usepackage[colorlinks = true,linkcolor = blue,urlcolor  = blue,citecolor = blue,anchorcolor = blue]{hyperref}

\usepackage{braket}
\newcommand{\ii}{{\rm i}}
\newcommand{\ee}{{\rm e}}
\newcommand{\dd}{{\rm d}}

\begin{document}

\title{Non-diffracting states in one-dimensional Floquet photonic topological insulators}

\author{Matthieu Bellec}
\email{bellec@unice.fr}\affiliation{Institut de Physique de Nice, Universit\'e C\^ote d'Azur, CNRS, 06100 Nice, France}
\author{Claire Michel}
\affiliation{Institut de Physique de Nice, Universit\'e C\^ote d'Azur, CNRS, 06100 Nice, France}
\author{Haisu Zhang}
\affiliation{Institute of Electronic Structure \& Laser, FORTH, Heraklion, Greece}
\author{Stelios Tzortzakis}
\affiliation{Institute of Electronic Structure \& Laser, FORTH, Heraklion, Greece}
\affiliation{Materials Science and Technology Department, University of Crete, 71003,
Heraklion, Greece}
\affiliation{Science Program, Texas A\&M University at Qatar, P.O. Box 23874 Doha,
Qatar}
\author{Pierre Delplace}
\email{pierre.delplace@ens-lyon.fr}\affiliation{Laboratoire de Physique, \'Ecole Normale Sup\'erieure de Lyon, Universit\'e de Lyon, Lyon, France}

\date{\today}

\begin{abstract}
One dimensional laser-written modulated photonic lattices are known to be particularly suitable for diffraction management purposes. Here, we address the connection between discrete non-diffracting states and topological properties in such devices through the experimental observation and identification of three classes of non-diffracting state. The first one corresponds to topologically protected edge states, recently predicted in Floquet topological insulators, while the second and third are both bulk modes. One of them testifies of a topological transition, although presenting topological features different from those of the edge states, whether the other one result from specific band structure engineering.
\end{abstract}

\maketitle

Manipulating the flow of light with unusual diffraction features has been enabled, during the last decades, by considering optical transport in structured photonic media~\cite{Joa08}. The analogy between solid-states physics and light propagation in specifically engineered arrayed structures allows the control of dispersion relations, which present in general a band structure, and thus of light transport properties.
In propagating geometries, where the propagation axis plays the role of time, optical devices can be envisaged both at short scales, using integrated photonic waveguide arrays~\cite{Chr03}, and at large scales, with multicore optical fibres~\cite{Rop11}.
In those so-called photonic lattices, the discrete diffraction, as opposed to continuous diffraction in homogeneous media, may exhibit uncommon behaviors as observed in various experimental realizations~\cite{Mor99, Eis00, Per02, Iwa05, Gar06, Lon06, Gar07, Sza08, Gar12}.
Importantly, when diffraction cancels, the associated non-diffracting states are of great importance since they offer the possibility to route information to specific regions on the lattice~\cite{Chr03, Per02a}.
In particular, a periodic modulation of the guides along the propagation axis was shown to exhibit striking non-diffracting modes that propagate with a well defined \textit{drift} angle~\cite{Dre13, Kar16}.

Interestingly, periodic modulations were also recently employed in the context of topological phases. While the concept of \textit{Floquet} topological insulators was first developed for irradiated semimetals and semi-conductors \cite{Oka09, Inoue10, Kitagawa11, Lindner11}, it found a spectacular experimental manifestation in out-of-equilibrium cold atom physics \cite{Jotzu14}, single-photon quantum walks \cite{Kit12} and photonic lattices \cite{Rechtsman2013, Mac17, Muk17}. In the latter system, a periodic modulation of the waveguides array along the propagation axis acts as a periodic driving. 
Topologically protected edge states may emerge in these driven systems and are non-diffracting by nature in photonic lattices.
This suggests a deep link between the existence of non-diffracting states, the longitudinal periodic modulation and the topological properties. 
For instance, do the modes reported in Ref.~\cite{Dre13, Kar16} possess a topological property? If so, are there non-diffracting modes in periodically modulated waveguide arrays that are not topological? 

In this paper, we answer these questions obersving and manipulating experimentally three kinds of non-diffracting modes in periodically modulated one-dimensional (1D) arrays of optical waveguides. Within the framework of the Floquet theory, we identify different mechanisms at the origin of these remarkable states. Two of them are found to be related to a topological property: i) the edge states, which emerge at the interface between two topologically distinct Floquet gapped phases, and ii) the \textit{drift} bulk states analogous to those observed in the optical beam rectification context~\cite{Dre13, Kar16}. The third ones propagate straight in the bulk and result from a flat dispersion relation of the Floquet spectrum.

\begin{figure}[t]
\centering
\includegraphics{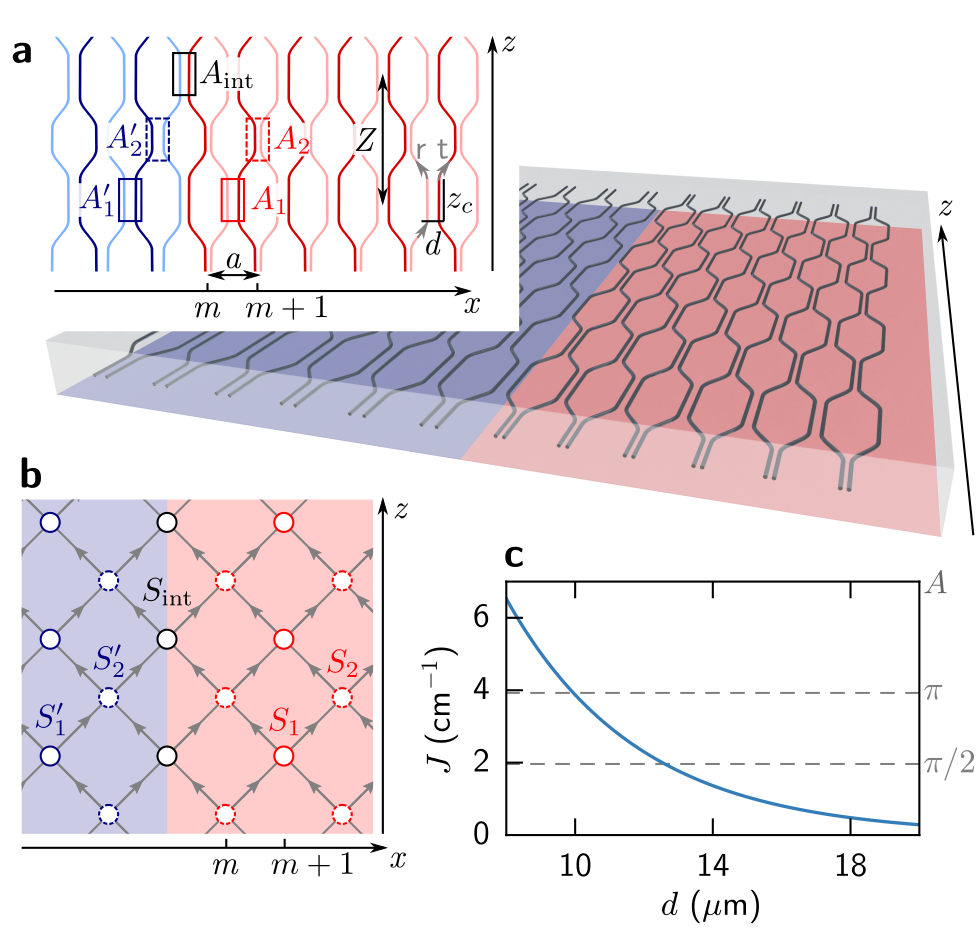}
\caption{(a) Sketch of the experimental realization of the 1D periodically modulated photonic lattice composed of two distinct arrays characterized by different phase-couplings $A_{1, 2}$ (red) and $A_{1, 2}'$ (blue). For each lattice, dark and light colors define the two sublattices separated by $a$ and $m$ labels the unit cells. $A_\text{int}$ is the phase-coupling at the interface. $Z$ is the period modulation. Gray arrows show how an incoming beam is transmitted ($t$) and reflected ($r$) by coupler defined by its separation $d$ and its length $z_c$. (b) Corresponding oriented network model where gray arrows are propagating links and colored nodes are scattering matrices $S_i$. (c) Exponential fit of the measured coupling $J$ at 633 nm versus waveguide separation $d$. Right axis scale corresponds to the phase-coupling $A=J z_c$.}
\label{Fig_Sketch} 
\end{figure}

Periodically modulated arrays of evanescently coupled waveguides offer the possibility to investigate light propagation in structures that combine both discrete and continuous periodicities. Consider the propagation of a scalar discrete optical field of the form $\exp[\ii  (k_x ma + k_z z)]$ through such a two-fold periodic array. The discrete periodic structure in the transverse direction yields a band structure for the wave vector $k_z(k_x)$, where the transverse quasi-momentum $k_x$ lives in a 1D Brillouin zone of length $2\pi/a$. In addition, the periodic modulation of period $Z$ of the guides along the propagation direction ($z$ axis in Fig.~\ref{Fig_Sketch}(a)) ensures a $2\pi/Z$ periodicity of $k_z$ itself. It follows that $k_z(k_x)$ displays a twofold periodic band structure analogous to the quasi-energy spectrum of periodically driven quantum systems~\cite{Sam73}. As presented in this paper, this striking property gives rise to various diffraction properties in the array.

\noindent The typical layout consists of a network of directional couplers as depicted in Fig.~\ref{Fig_Sketch}(a). We fabricate this structure in a 10 cm long fused silica sample (Suprasil 311, Hereaus) using the second harmonic output (515 nm) of a Yb:KGW regenerative laser system (Pharos, Light Conversion) delivering 150 nJ pulses with 190 fs duration at 200 kHz repetition rate. The laser beam is tightly focused 200 $\mu$m inside the glass using a 20$\times$ microscope objective with NA=0.40. The sample is moved with high-precision translation stages (Aerotech ANT series) at 0.5 mm/s. Each fabricated waveguide supports a single mode propagation at 633 nm. By measuring the output intensity of directional couplers for various waveguide separations $d$, we obtained the coupling strength $J$ which takes into account the extra coupling from the waveguide bending. The corresponding exponential fit is shown in Fig.~\ref{Fig_Sketch}(c). For the experiments, the whole light propagation in the array is monitored by laterally visualizing the visible fluorescence excited at 633 nm and emitted by the color centers created during the waveguide fabrication.

\noindent As shown in Fig.~\ref{Fig_Sketch}(b), our photonic lattice can be seen as an oriented network with propagating links (gray arrows) and scattering events (colored circles). It is worth noting that the signal is driven from bottom to top, unlike other two-dimensional oriented lattices in which the signal travels in both directions and whose topological properties have been recently investigated \cite{PasekChong14, HuPRX15, tauber_delplace_NJP15, delplace16}. In this configuration, the scattering at each node is captured by a $2\times 2$ unitary matrix whose coefficients describe how light in an incoming waveguide is reflected and transmitted into the two outcoming waveguides. Indeed the propagating and bending losses being global, they decouple from the scattering processes.     
Considering first one of the two lattices (e.g. the red one), one distinguishes two scattering regions where guides are designed with distinct separations $d_1$ and $d_2$  and thus distinct coupling strengths $J_1$ and $J_2$ (Fig.~\ref{Fig_Sketch}(c)). 
With fixed couplers of length $z_c =11.88$ mm, the scattering coefficients are ruled by the phase $A_i = J_i z_c$ with $i=1,2$ (hereafter the phase-coupling). One can then assign a scattering matrix $S_i$  to each coupler, and express it as a function of the phase coupling as
\begin{equation}
S_i = \begin{pmatrix}
\cos{A_i} & -\ii \sin{A_i}\\
-\ii \sin{A_i} & \cos{A_i}
\end{pmatrix} \, .
\label{eq_scattering}
\end{equation}
\begin{figure}[t]
\centering
\includegraphics{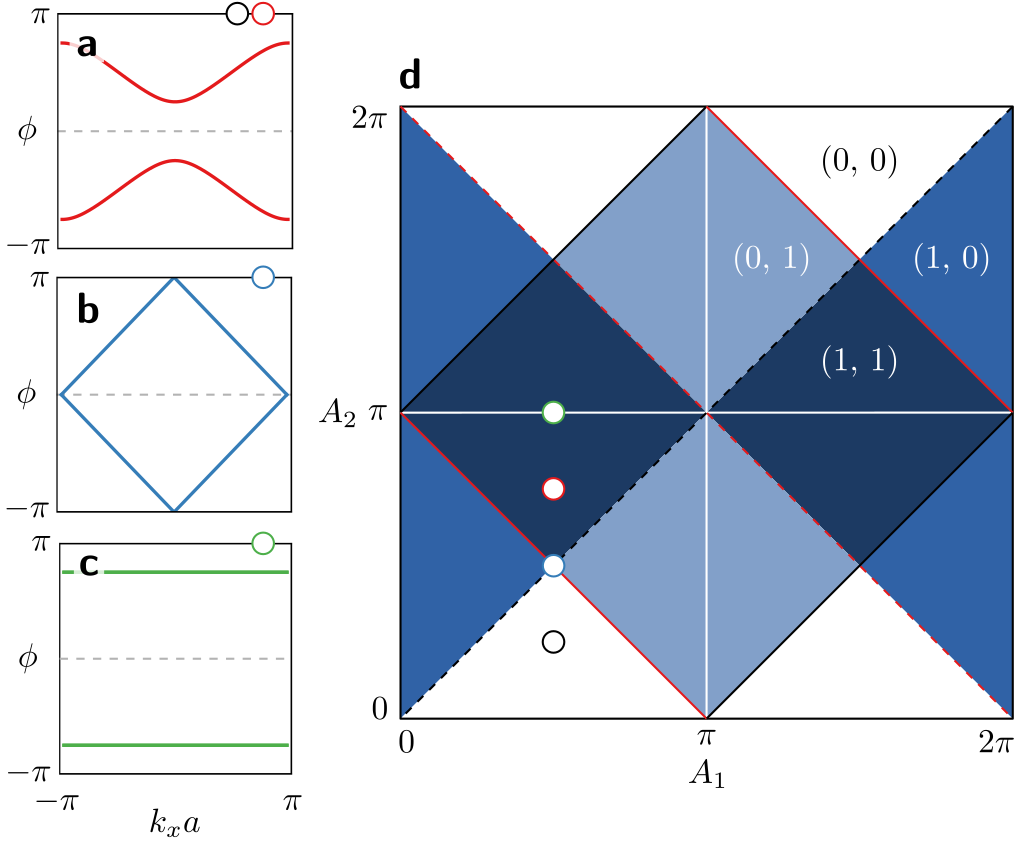}
\caption{Dispersion relations (a--c) and phase diagram (d) calculated for infinite systems with $A_1$ and $A_2$ ranging from 0 to $2\pi$. (a) Typical two-bands dispersion. Gaps closures occur at $k_z Z=0$ [dashed line in (d)] and $k_z Z=\pi$ [full line in (d)] for at $k_x=0$ [red in (d)] or $k_x=\pi$ [black in (d)]. (b) Linear dispersion at bi-critical points [full and dashed diagonal lines crossings in (d)]. (c) Flat bands dispersion appearing at critical points [white line in (d)]. (d) Colors correspond to $(\nu_0, \nu_\pi) = (0, 0)$ (white), $(0, 1)$ (light blue), $(1, 0)$ (medium blue) and $(1, 1)$ (dark blue).}
\label{Fig_Diagram}
\end{figure}

Light propagation in the bulk is studied by considering the scattering over a period $Z$ of the light field wavefunction $\boldsymbol{\psi}(z)$. This is encoded by the evolution operator $U_0$ defined as $\boldsymbol{\psi}(z+Z) = U_0 \boldsymbol{\psi}(z)$. The bulk optical field $\boldsymbol{\psi}$ reads in the Bloch basis as a two-component wavefunction resulting from the two guides of the unit cell (dark and light in the inset of Fig.~\ref{Fig_Sketch}(a)). 
Then the evolution operator $U_0=U_0(k_x)$ consists of a sequence of scattering matrices as 
\begin{equation}
U_0(k_x) = B^\dagger(k_x) S_2 B(k_x) S_1
\label{eq_U0}	
\end{equation}
where $B(k_x)$ is a unitary matrix taking into account the Bloch phase $\exp(\ii \, k_x a)$ accumulated when waveguides from different unit cells couple. Eqs.~(\ref{eq_scattering}, \ref{eq_U0}) are obtained by computing the evolution operator within a stepwise time dependent tight-binding model. After length $Z$ has been spanned, each $\boldsymbol{\psi}$ component has accumulated the same phase $\phi=k_z Z$ so that the light field satisfies the eigenvalue equation
\begin{equation}
U_0(k_x) \, \boldsymbol{\psi}(k_x) = \ee^{\ii \phi(k_x)} \, \boldsymbol{\psi}(k_x) \, .
\label{eq_heuristic_equation}
\end{equation}
A direct diagonalisation of $U_0(k_x)$ shows that the two solutions $\phi_\pm(k_x)=k_z^{\pm}(k_x)Z$ consist of two bands which generically do not touch for any $k_x$ in the 1D Brillouin zone (Fig.~\ref{Fig_Diagram}(a)).

\noindent Such gapped Floquet systems can display non trivial topological properties that differ from those of topological insulators at equilibrium. Indeed, they can develop \textit{anomalous} topologically protected boundary states while all the topological invariants defined for the \textit{bands} vanish. This requires to define new topological indices that correctly account for the full periodic evolution \cite{Rudner13, Carpentier15, Fruchart16, delplace16}.
In particular, depending on both the dimension of the system and  its symmetries, a bulk topological index $\nu_\kappa$ can be assigned to each \textit{gap}, labelled by $\kappa$, rather than to a band \cite{Rudner13, Carpentier15, CarpentierNPB15, Fruchart16, delplace16}. This topological index is directly related to the existence (and number) of protected boundary states in the gap $\kappa$ in finite geometry.
As long as there exists a symmetry axis $z\rightarrow -z$ of the lattice of Fig.~\ref{Fig_Sketch}(a) with respect to some origin, the operator $U_0(k_x)$ holds a chiral symmetry.
Following previous theoretical works~\cite{Asboth_Tarasinski_Delplace, Fruchart16}, this allows us to define a bulk topological index $\nu_\kappa$ for each of the two gaps $\kappa=0$ and $\kappa=\pi$ (see Fig.~\ref{Fig_Diagram}(a)).
Four distinct topological phases, characterized by different values of the couple ($\nu_0,\nu_\pi$), are found when varying the phase-couplings $A_1$ and $A_2$ as represented in the phase diagram of Fig.~\ref{Fig_Diagram}(d). Note that it is similar to a previous study for a single-photon version of the problem, but where a different topological characterization was proposed~\cite{Kit12}. 

We now consider a system with two different bulk properties which are separated by an interface. When the difference between the bulk topological indexes of each side $\nu_\kappa-\nu'_\kappa$ does not vanish, interface states  are expected to emerge in the gaps $\kappa=0$ or $\kappa=\pi$, meaning that they carry a quantized phase $\phi_\kappa=\kappa$  when propagating over a distance $Z$. In particular, for $\nu_0-\nu'_0 = \nu_\pi-\nu'_\pi = 1$, both $0$ and $\pi$-phase anomalous modes are expected.
In order to practically investigate such states, we now consider two finite chains of waveguides. The two arrays are characterized by a set of two phase-couplings ($A_1,A_2$) and ($A'_1,A'_2$) (red and blue parts in Fig.~\ref{Fig_Sketch} respectively), such that their topological invariants, $(\nu_0, \nu_\pi)$ and $(\nu_0', \nu_\pi')$, can be different. The last waveguide of the blue chain is coupled to the first waveguide of the red chain by a phase-coupling $A_{\text{int}}$ which defines an interface along $z$. 
\begin{figure}[t]
\centering
\includegraphics{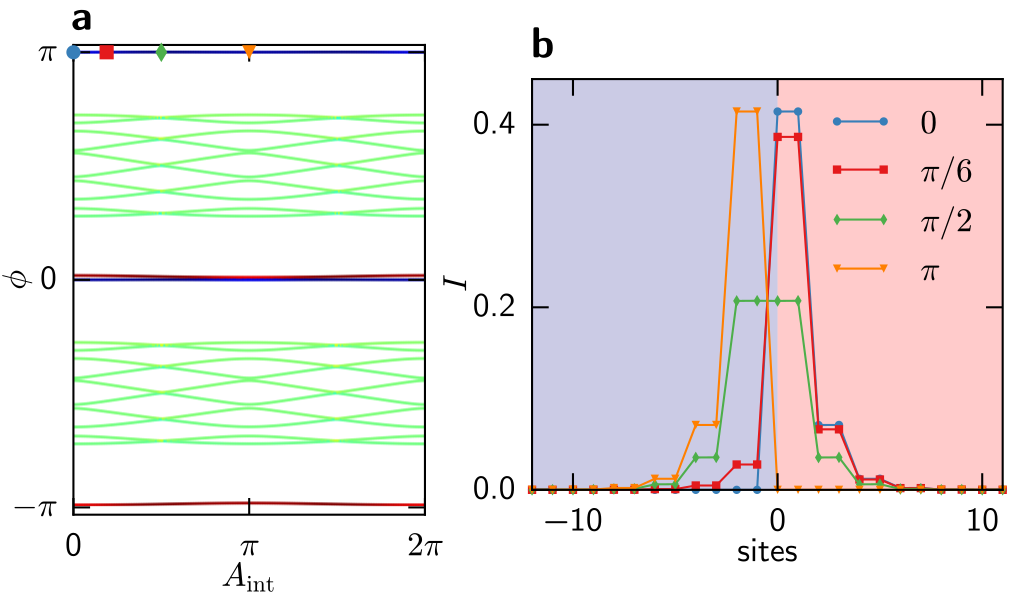}
\caption{(a) Eigen-phases $\phi=k_z Z$ of two arrays of $12$ waveguides in a cylindrical geometry as a function of the interface phase-coupling $A_{\text{int}}$. The two arrays are characterized by $(A_1, A_2) = (\pi/2, \pi/4)$ and $(A_1', A_2') = (\pi/2, 3\pi/4)$ [respectively black and red circles in Fig.~\ref{Fig_Diagram}(d)], corresponding to $\nu_\pi-\nu'_\pi = \nu_0-\nu'_0 = 1$. Green corresponds to bulk states, blue and red to localized states at the interfaces. For clarity, the degeneracy is lifted by adding a small potential on one of the interfaces. (b) Intensity of the corresponding $\pi$-phase modes for various $A_{\text{int}}$.}
 \label{fig_interface_mode} 
\end{figure}
We consider here two arrays characterized by the phase-couplings $(A_1, A_2) = (\pi/2, \pi/4)$ and $(A_1', A_2') = (\pi/2, 3\pi/4)$ corresponding to black and red circles in Fig.~\ref{Fig_Diagram}(d). 
Numerical calculations of the phase spectrum are shown in Fig.~\ref{fig_interface_mode}(a).

\begin{figure*}[t!]
\includegraphics{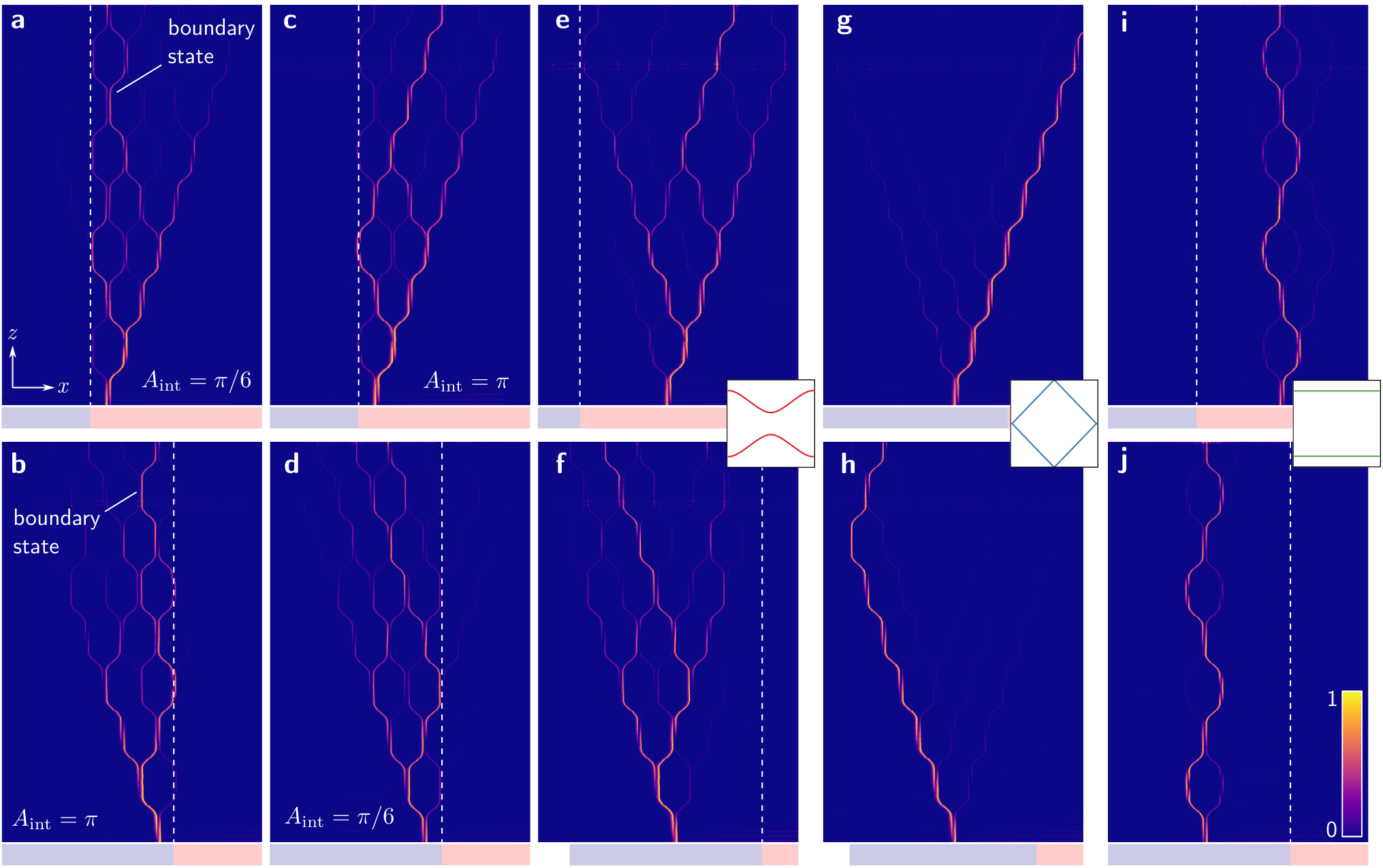}
\caption{Experimental images of the light propagation for a single waveguide excitation in the dark (resp. light) sublattice [top (resp. bottom) panels]. The image size is 100 mm $\times$ 1.1 mm. The dashed line delimits the interface between red and blue arrays. (a--f). Lattice parameters : $(A_1, A_2) = (\pi/2, \pi/4)$ and $(A_1', A_2') = (\pi/2, 3\pi/4)$ [respectively black and red circles in Fig.~\ref{Fig_Diagram}(d)], corresponding to $\nu_0-\nu'_0 = \nu_\pi-\nu'_\pi = 1$, with $A_{\text{int}} = \pi/6$ (a,d), $\pi$ (b,c). (e,f). Bulk excitation. (g,h). Lattice parameters : $(A_1, A_2) = (A_1', A_2') = \pi/2$ [green circle in Fig.~\ref{Fig_Diagram}(d)] with $A_{\text{int}} = \pi/2$. (i,j). Lattice parameters : $(A_1, A_2) = (A_1', A_2') = (\pi/2, \pi)$ [blue circle in Fig.~\ref{Fig_Diagram}(d)] with $A_{\text{int}} = \pi$.}
\label{Fig_Exp}
\end{figure*}

The experimental setup (Fig.~\ref{Fig_Exp}(a)) presents two boundaries in addition to the interface between the two arrays.
To get rid of the additional localized states that may appear at these edges, we experimentally built large enough arrays and we numerically coupled their two extremities with the same $A_{\text{int}}$.
The calculation of the boundary modes shows that their existence is independent of the coupling $A_{\text{int}}$ between the two arrays, which illustrates their topological robustness. However, as shown in Fig.~\ref{fig_interface_mode}(b), their intensity profile oscillate from one side of the interface to the other when tuning $A_{\text{int}}$.

Figure~\ref{Fig_Exp} shows the corresponding experimental images of the light intensity propagation for a single waveguide excitation at the vicinity of the interface (a--d), delimited by the dashed line, and in the bulk (e,f). In Fig.~\ref{Fig_Exp}(a) [resp. (b)], we clearly observe a boundary state in the red (resp. blue) array for $A_{\text{int}} = \pi/6$ (resp. $\pi$). Although not predominant, the local excitation of only one waveguide necessarily excites bulk modes, with a non-vanishing relative weight. The comparison of Fig.~\ref{Fig_Exp}(a) [resp. (b)] with Fig.~\ref{Fig_Exp}(e) [resp. (f)] shows that the bulk mode is indeed qualitatively visible.
Note that $0$ and $\pi$-modes are degenerated in intensity and only differ with their phase profiles. In the actual configuration, both modes are excited and additional (ongoing) experiments are required to discriminate them.  
On the contrary, as shown in Fig.~\ref{Fig_Exp}(d) [resp. (c)], for $A_{\text{int}} = \pi/6$ (resp. $\pi$), when excited in the blue (resp. red) array, only bulk mode propagate. Here the comparison with the bulk modes excitation in Fig.~\ref{Fig_Exp}(e-f) is much more obvious. 

These results present an experimental observation of anomalous boundary states in a 1D Floquet photonic topological insulator. This is the first class of modes mentioned in the introduction which establishes a clear link between the existence of topologically protected edge states and diffractionless propagation.

Besides, the phase diagram in Fig.~\ref{Fig_Diagram}(d) shows bi-critical points (at full and dashed diagonal lines crossings) between gapped phases with distinct bulk topological invariants. At these points, given by ($A_1, A_2$) = ($\frac{\pi}{2}(2p+1),\frac{\pi}{2}(2p'+1)$), the two gaps close simultaneously leading to degeneracy points at $\phi =0$ and $\phi=\pi$ (Fig.~\ref{Fig_Diagram}(b)). Importantly, these transition points are accompanied by an additional sublattice symmetry: the evolution operator $U_0$ becomes diagonal and thus commutes with $\sigma_z$, which is not true in general. It follows that the two bulk modes belong to opposite sublattices and remain uncoupled while carrying opposite group velocities in the transverse direction. As a result, the excitation of an arbitrary waveguide necessarily always coincides with an eigenmode of the system as shown experimentally in Fig.~\ref{Fig_Exp}(g,h). 
This is a remarkable property of the bi-critical points, since it generates diffractionless bulk states with a transverse \textit{drift} angle whose sign is reversed when changing the sublattice to which the excited guide belongs. Note that similar states have been found numerically~\cite{Kar16} and observed experimentally~\cite{Dre13} in modulated waveguides arrays. They were interpreted in terms of optical beam rectification. This interpretation is not inconsistent with our results but we go further by identifying them as a signature of a critical gapless Floquet phase.

\noindent This striking behavior is independent of the specific excited site, once the sublattice is fixed. Moreover, the excitation of a single waveguide corresponds to a global probe in quasi-momentum space. This suggests that these drift diffractionless states may reflect another topological property of the system. Clearly the existence of these states lies on the  periodicity of the phase spectrum in $k_z Z$. They  are thus specific to unitary systems.
This allows us to define the winding number 
\begin{equation}
w_{\pm} = \frac{1}{2\ii \pi} \int_{-\pi/a}^{\pi/a} \dd k_x  \bra{\psi_\pm} U_0^{\dagger} \left( \partial_{k_x} U_0 \right) \ket{\psi_\pm}
\end{equation}
of the Floquet bulk state $\psi_\pm$ which is non-zero at the gapless transition points. 
This winding number, that can be rewritten as a first Chern number, is proportional to the average displacement in the transverse direction, over a period $Z$ \cite{KitagawaPRB10}. This leads to a quantized transversed displacement, that is clearly observed in Fig.~\ref{Fig_Exp}(g,h)
This is known as a topological pumping process, and usually arises in 1D equilibrium gapped phase adiabatically modulated in time (or space) \cite{Thouless1983, KitagawaPRB10, Kra12, Lohse15}. However, here, the periodic modulation is specifically non-adiabatic, since the frequency driving $2\pi/Z \approx \pi$ cm$^{-1}$ is larger than the typical coupling amplitude $J\approx 1$ cm$^{-1}$, or equivalently, because the drive period is smaller than the coupling length. Besides, in the absence of time-reversal symmetry breaking, the Chern number vanishes, which is consistent with the fact that the winding of the two Floquet bands compensate each other, i.e. $w_-+w_+=0$. However, as explained above, the two different branches of the spectrum can be excited separately so that light propagates without diffracting in one direction only. This artificially breaks time-reversal symmetry at the input of the array, similarly to what is performed e.g. in two-dimensional coupled resonators optical waveguides arrays to simulate an optical analog of a quantum Hall phase \cite{Haf11}.

Finally, the third kind of diffractionless mode can be excited e.g. at the interface of two lattices that carry the same bulk topological index. However, they do not benefit from any topological protection inherited from chiral symmetry: their existence depends on the phase-coupling $A_\text{int}$ at the interface, unlike what is shown in Fig.~\ref{fig_interface_mode}(a). In particular, they are maximally localized at the edge and possess a Floquet phase of exactly $\kappa$ when the transmission coefficient with one of the two adjacent waveguides vanishes, that is for either $A_1=p\pi$ or $A_2=p\pi$, with $p$ an integer [white lines in Fig.~\ref{Fig_Diagram}(d)]. This is actually not specific to boundary modes, and can be engineered in the bulk. There, it actually corresponds to the case where the two Floquet bands are flat, as shown in Fig.~\ref{Fig_Diagram}(c). 
As a consequence, the degree of diffraction $\partial^2 k_z /\partial{k_x^2}$ vanishes for every $k_x$. This is remarkable as it involves a non-diffracting behavior of the light field for any excitation of the array, and not only for wave packets centered around specific $k_x$~\cite{Eis00}. In particular, diffraction vanishes for single-guide excitations, that are not  eigenmodes of the system in contrast with the drift diffractionless bulk modes. This behavior is shown experimentally in Fig.~\ref{Fig_Exp}(i,j) where \textit{straight} diffractionless bulk modes are indeed observed for various positions of the excitation.

To summarize, we have observed three kinds of non-diffracting modes in a 1D array of evanescently coupled optical waveguides. First the edge states, that necessarily emerge at the interface between two arrays whose scattering matrices, ruling the evolution of the optical field along the propagation axis, carry distinct bulk topological indices. Second the drift bulk modes, that arise at the double transitions of the Floquet phase diagram by restoring a sublattice symmetry. Third the straight bulk modes, that result from a flat dispersion relation of the Floquet spectrum. 
It worth noticing that while the drift modes can be understood as the manifestation of a (transverse) topological pumping, the straight modes can instead be seen as the consequence of a dynamical (transverse) localization~\cite{Gar12}. 
Furthermore, while they are of different origin, both the drift modes and the straight modes are insensitive of the way to excite the array. However, their existence requires some fine-tuning of the phase couplings $A_i$. In contrast, the edge states, which are more tricky to excite precisely, do not require any fine-tuning of the couplings, neither in the bulk nor at the interface.

\begin{acknowledgments} 
P.D. thanks J. Li and M. Fruchart for fruitful discussions. M.B. and P.D. thank J.M. Jarre for his support. This work was supported by the French Agence Nationale de la Recherche (ANR) under grant TopoDyn (ANR-14-ACHN-0031). H.Z. acknowledges support from the EU FP7-REGPOT-2012-2013-1 (no 316165).
\end{acknowledgments}

\bibliography{biblio}

\end{document}